# Cosmic Rays and the Evolution of Earths Climate During the Last 4.6 Billion Years[a]

**Henrik Svensmark**

Danish Space Research Institute, Juliane Maries Vej 30, 2100 Copenhagen Ø,

Denmark

Variations in the flux of Galactic Cosmic Rays (GCR) at Earth during the last 4.6 billion years are constructed from information about the Star Formation Rate (SFR) in the Milky Way and the evolution of solar activity. The variations of GCR show a remarkable resemblance to changes in Earth's climate during the period considered, suggesting that Earths climate is closely linked to the evolution of our Milky Way. The link could be significant in the solution of the "faint sun climate paradox".

98.35.Hj, 92.40.Cy, 92.70.Gt, 98.70.Sa

Quite surprisingly, new results involving solar induced variations in atmospheric ionization by Galactic Cosmic Rays (GCR) suggest that they are important in climate change[1]. It has been shown that low-level clouds seem to be responding to solar cycle variations in GCR and so influence the energy budget of the Earth[2,3,4]. Related, but on much longer timescales, new work has shown a remarkable correlation between variations in the source of GCR caused by the solar systems passage through the spiral arms of our Milky Way, and variations in Earth's climate during the last 500 million years[5,6]. This paper takes this idea one step further by reconstructing variations in the GCR flux on even longer timescales ranging from 400 million years to 4.6 billion years, by considering variations in the galactic source of GCR and the evolution of solar activity. It is shown, that even on long timescales Earth's climate varies with the reconstructed GCR flux. If the relation is real, it suggests that the whole history of the Milky Way influences the evolution of climate. In addition such a relation could be a significant contribution in the solution of the paradox of the young faint sun, i.e. the contrast of a weak young sun and warm early Earth.

---

[a] Submitted to Physical Review Letters 16th June 2003

The paper is organized as follows: First the solar modulation of GCR at 1 AU is estimated. Then the variations of the GCR-source are estimated based on the star formation rate (SFR) in the Milky Way. Finally, the GCR flux at 1 AU modulated both by solar effects and changes in GCR source is constructed for the history of the solar system and is compared with the Earth's climate history.

GCR are energetic particles (GeV, mainly protons) that are accelerated in shock fronts of supernova (SN) explosions. Before GCR can reach the Earth they must penetrate the Heliosphere, the region of space dominated by the outflow of magnetized solar wind from the Sun and extends far beyond all the planets out to a distance of about 100 - 150 AU. The solar wind is carrying the Sun's magnetic field and is so capable of modulating the charged flux of GCR in the solar system. Solar modulation of cosmic rays reaching the Earth (1 AU) is described by a transport differential equation first posed by Parker[7]. Neglecting drifts, and only considering high-energy (GeV) cosmic rays, one obtains the widely used force-field approximation[8], which corresponds to charged particles in a potential $\Phi = (R-1) V/3\kappa_0$, where $V$ is the solar wind velocity, $\kappa_0$ is the GCR diffusion constant induced by magnetic scattering, and $R$ is the size of the heliosphere (in AU). In this approximation the GCR differential particle intensity at 1 AU, $J(T)$, is related to the local interstellar GCR differential particle intensity at R as

$J(T) = J_R(T+\Phi) \ [T(T+2T_0)]/[(T+\Phi)(T+\Phi+2T_0)]$, where $T$ is the kinetic energy, and $T_0$ is the rest energy of the particle (see e.g. Borella et al.[9]). Changes in solar activity with time are accounted for by $\Phi = \Phi(t)$ in the above equation. The temporal variations in the interstellar GCR differential particle intensity is approximated by $J_R(T+\Phi(t)) \propto n_{GCR}(t) \ (T+\Phi(t)+T_0)^{-2.7}$, where $n_{GCR}(t)$ is a time varying number density. Inserting $J_R$ into the expression for $J$, with $\Phi = \Phi(t)$, one obtains the GCR differential particle intensity relative to present day (t=0) as,

$$j(T,t) = \frac{J(T,t)}{J(T,0)} = \frac{n_{GCR}(t)}{n_0} \left(\frac{T+T_0+\Phi_0}{T+T_0+\Phi(t)}\right)^{2.7} \left(\frac{(T+\Phi_0) \ (T+\Phi_0+2T_0)}{(T+\Phi(t)) \ (T+\Phi(t)+2T_0)}\right),$$

where $\Phi(0) = \Phi_0$, and $n_{GCR}(0) = n_0$. This approximation is valid for energies larger than few hundred MeV, and will be used here for GeV particles. Estimates of $\Phi_0$ range from 0.2 –1.2 GV. Here the medium value $\Phi_0 = 0.75$ is adopted[9].

The task is now to determine the functions $\Phi(t)$, and $n_{GCR}(t)$. Insight to the evolution of our Sun is gained from extensive studies of solar proxies with ages from 100 Myr to 10 Gyr[10]. The Sun young was rotating at a rate at least 10 times faster than today. As a consequence, the Sun had a vigorous magnetic activity with coronal X-ray and EUV emissions up to thousand times stronger than today. In addition the sun had a denser solar wind[10,11]. These variations could be included in the force-field potential $\Phi = (R-1) \ V/3\kappa_0$ as variations in the solar wind velocity, the diffusion constant, and the size of the heliosphere. However, for simplicity the focus will be on variations in the heliospheric radius. Reasons are: 1) The derived force-field potentials based on $R$ will in itself cover a large range that in principle could contain uncertainty in all of its parameters. 2) There are no clear observational evidence for systematic variations $V$ or in $\kappa_0$ caused by the solar cycle (se e.g. Jokipii in[10]).

A pressure balance between the solar wind ram pressure $\dot{M}V/R^2$ and $P_{LIM}$ the pressure of the local interstellar media gives the heliospheric radius. $\dot{M}$ is the solar mass loss rate. Assuming that the interstellar pressure is constant (on timescales ~ 0.4 Gyr), the temporal variation in the distance to the heliopause is given by[11]

$R(t) = \sqrt{V\dot{M}(t)/P_{ISM}} = \sqrt{const * \dot{M}(t)}$. Inserting $R(t)$ in the force-field potential one derives,

$$\Phi(t) = \Phi_0 \frac{(\sqrt{const * \dot{M}(t)} - 1)}{(\sqrt{const * \dot{M}_0} - 1)} \approx \Phi_0 \sqrt{\dot{M}(t)/\dot{M}_0} \quad .$$

Where $\dot{M}(0) = \dot{M}_0$ is the present average solar mass loss rate. The function $\dot{M}(t)$ has been estimated from solar like stars. One study[11] finds a mass loss rates corresponding to $\dot{M}(t) \propto \exp\left[-(2.53 \pm 0.51) * \sqrt{t}\right]$ (based on Walter & Berry in[10]), or to $\dot{M}(t) \propto t^{-2.00 \pm 0.52}$.

Figure 1 shows $\Phi(t)/\Phi_0$ for these two cases. The error bars are given by the uncertainty in the exponents. Notice that the two estimates are similar, and that the young Sun's force-field potential was 10 – 100 times stronger than today. Now $\Phi_0, \Phi(t)$ and $n_{GCR}(t)/n_0 = 1$, can be inserted into $j(T,t)$. Figure 2 shows the modulation of GCR particles with energies 5, 10, 20 and 30 GeV, using $\Phi(t) \propto \left(\exp\left[-(3.04) * \sqrt{t}\right]\right)^{0.5}$ as the medium force-field potential. It is seen that the modulation by the young Sun was severe for lower GCR energies, nearly depleting the inner heliosphere of these particles.

Stars that end their life in SN (type SN II and SN Ib) explosions have large masses (initial mass 10 times larger that our sun) and relative short life times (5-100 million years) [12]. Variations in the SFR also reflect the birth rate of such massive stars, and with their short lifetimes, also the rate of SN explosions. As SN's are the source of GCR's, the GCR- source can be described as proportional to SFR. Other types of SN explosions are not relevant since their rate explosion rate is much lower[12]. Figure 3, adapted from Rocha-Pinto et al.[13], shows the temporal SFR history of the Milky Way in time steps of 0.4 Gyr (solid curve), or as interpreted here, the history of GCR intensity variations $n_{GCR}(t)$ in the interstellar media. It is important to note that the variations in GCR due to spiral arm crossings are on a shorter time scale ~0.140 Gyr than the 0.4 Gyr time step used here. Therefore the GCR shown in figure 3 represent an average Milky Way interstellar GCR flux over the entire orbit of the solar system around the Galactic center (one revolution for the solar system last about 0.24 Gyr). The data in the figure has been normalized to the average SFR (or GCR flux) over the last 15 Gyr. The dashed line is obtained by disregarding outliers beyond 2σ shown as a dotted line[13]. This indicates first of all that the general features of the SFR or GCR history are robust. There is however another reason to exclude the outliers. They represent, in the case of the most resent bins, very close stars that have not had time to disperse in the galactic disk. By excluding these outliers a better global SFR signal for the most resent bins is expected.

Finally, $\Phi_0$, $\Phi(t)$ and $n_{GCR}(t)$ can be combined in $j(T,t)$ to give an estimate of the modulation of GCR due to both solar evolution and varying interstellar GCR flux. The result for 10 GeV GCR particles is shown in figure 4 (top panel) and is the main result in this paper. There are a few general features that should be noted in this figure. (1) In the beginning of the solar systems history the solar magnetic activity was so high that it very effectively screened out or severely damped the flux of GCR. (2) The GCR flux increases towards a local maximum around 2.2 Gyr before present. (3) Thereafter follows a local minimum around 1.5 Gyr before present, and finally (4), a maximum in GCR during the last 1 Gyr. These features are fairly robust, and do not depend on which form of force-field potential that is used, or the exact energy of the GCR particles. Comparing figure 3 and the top panel of figure 4 it is seen that the variations in the intensity $n_{GCR}(t)$ of in the early solar system are not very important due to the strong damping of GCR by the active young Sun.

The lower panel of figure 4 contains two attempts to reconstruct Earths climate during the history of the solar system. The solid curve shows the relative variation in sea level, and the dashed curve is the relative variations in atmospheric $CO_2$ concentrations[14]. The curves are a schematic way of indicating variations in climate. They are based on a vast amount of irregularly distributes data[15,16]. The extremes indicate either a Greenhouse climate (Warm) or an Icehouse (Cold), corresponding to high sealevel/$CO_2$ level or low sealevel/$CO_2$ level respectively. The gray areas are known periods with glacial periods, i.e. cold periods[15,16]. It is remarkable that no glaciations are found prior to 2.7 Gyr, and no glaciations are found between 2.2 Gyr and 1.0 Gyr. Comparing the two panels in figure 4 one sees a remarkable agreement in the variations in GCR and Earth's climate. It has been suggested that low cloud cover increases with GCR intensity, leading to a cooling, and vice versa[2-4]. Shaviv also noticed that some features of the SFR variation resembled features of climate[6].

Although there is a general agreement between the GCR flux and climate, it is not suggested that GCR is the only influence of importance. None the less, it might be of significance. According to the standard model of stellar evolution the young Sun was fainter by about 30% compared to present the days value. As a consequence (assuming present day atmospheric composition) the surface temperature would be below waters freezing point prior to 2 Gyr before present[17]. However, biological and geological records dating back as far as 3.8 Gyr indicate a warm climate with liquid water and sediments showing a biological activity that is much higher than what would be expected if the Earth were frozen. There are other indications of a much warmer early climate that present, for example high ocean temperatures ~40 $^0$C in the period 2.6-3.5 Gyr before present[18]. The contrast of a weak young sun and a warm young Earth is usually called "the Faint Sun Paradox", and has been the subject of many studies. Even for the planet Mars there are indications of a warm past with liquid oceans ~ 3.8 Gyr before present[19]. One attempt to resolve this paradox is by assuming a stronger greenhouse effect than the present one. However, it is not obvious that a strong greenhouse effect is the solution[20]. Another possibility could be a brighter young sun, i.e. a deviation from the standard model of solar evolution. A consequence of having higher mass loses in the beginning is that the initial mass of the sun was significant higher leading to a brighter young sun[21]. There are however constraints from helioseismology on how massive the young sun could have been[20], secondly indirect observations of mass loses of solar like stars seems to indicate mass losses for the sun insufficient to explain the faint sun paradox for Mars and only marginally for Earth[22].

GCR influence on climate can therefore be an important part in resolving the Faint Sun Paradox for both Earth and Mars, for instance through low cloud cover modulation mentioned above. The exact size of the influence can't be estimated at present. GCR variation caused spiral arm crossings and associated variations in ocean temperatures discussed by Shaviv and Veizer[23], indicate variations of the order ~ 5 $^0$C, and since GCR modulation by the young sun is even stronger, variations could be of the order ~10 $^0$C. However, one should note that solar modulation changes the spectral composition of GCR at 1 AU, while SFR (or spiral arm crossings) retains the spectrum.

A basic microphysical mechanism responsible for a link between GCR and climate has not been identified. There have been several suggestions involving clouds[24], but at the moments it is unclear. If one mechanism is active over most of history of the Earth it operates even though the atmospheric compositions has been changing. The observed correlation between GCR ionization and atmospheric properties, ranging from years to billion of years, warrants a thorough investigation.

In any case it is a fascinating thought that the history of the whole galaxy has influences the evolution of climate and life on Earth, and will continue so in the future.

Acknowledgement – thanks to T. Neubert, N. Marsh for helpful discussions, and thanks to the Carlsberg foundation for support.

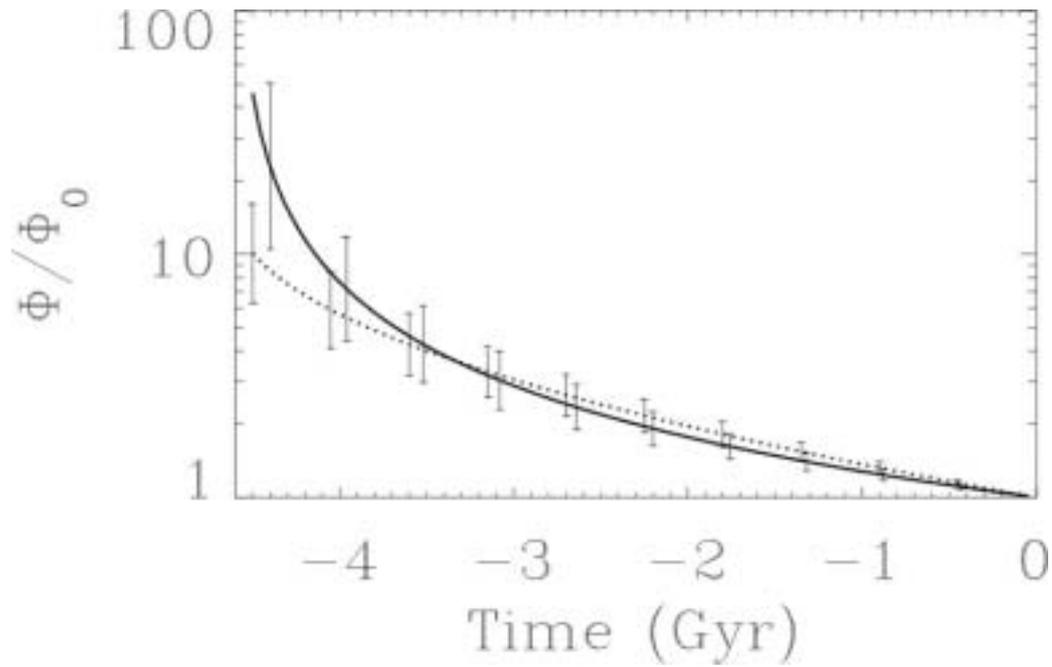

Fig 1. Estimated variations in the force-field potential due to the evolving solar activity is shown. The various solid curves are based on estimated solar wind mass-loss (see text).

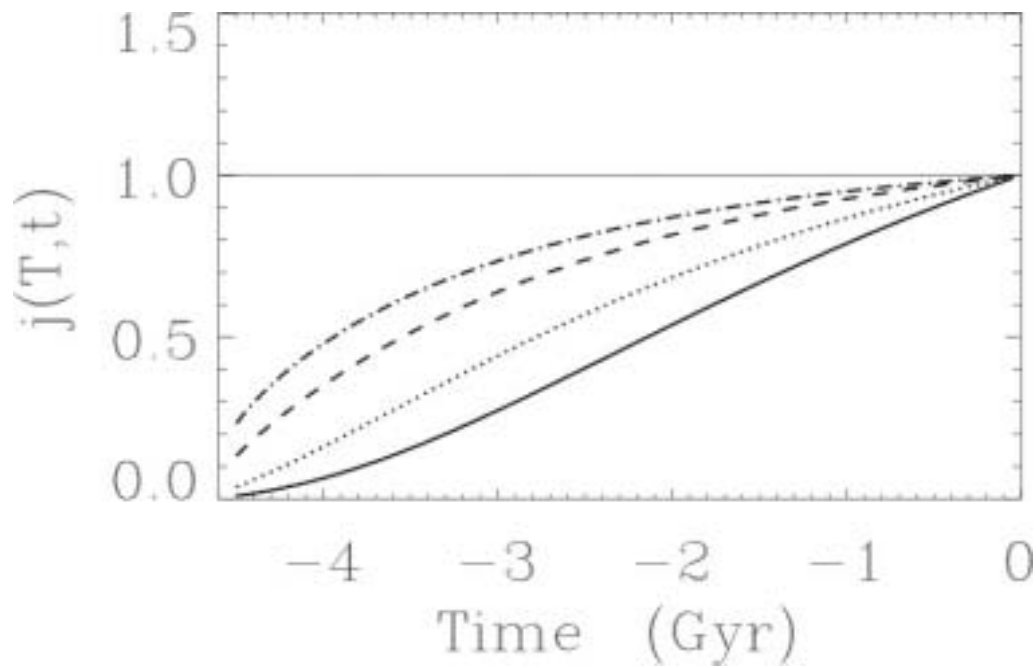

Fig 2. The modulation of cosmic rays at 1 AU as a function of time normalized to the present day intensity of GCR. The solid curve is for 5 GeV, doted curve for 10 GeV, dashed curve for 20 GeV and dashed-dotted curve 30 GeV.

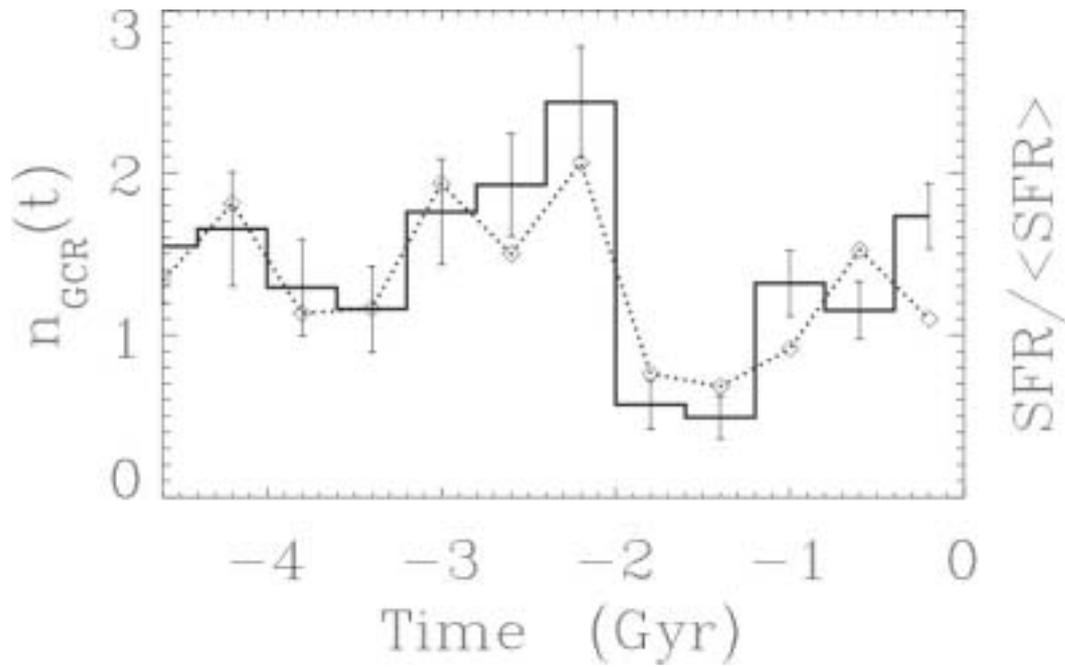

Fig 3. The histogram indicates the star formation rate (SFR) in the Milky Way over the history of the Solar System (the last 4.6 Gyr) adapted from Rocha-Pinto et al. The figure can also be interpreted as showing the variation in Galactic Cosmic Rays (GCR) (see text). SFR and GCR are normalized to the average <SFR> or <GCR> in the Milky Way over the last 15 Gyr. The temporal resolution is 0.4 Gyr. The Error bars are Poisson counting uncertainty. By discarding 2σ outliers in part of the transformation that leads to SFR, the dashed line is obtained. This indicates that the general features of SFR construction are robust.

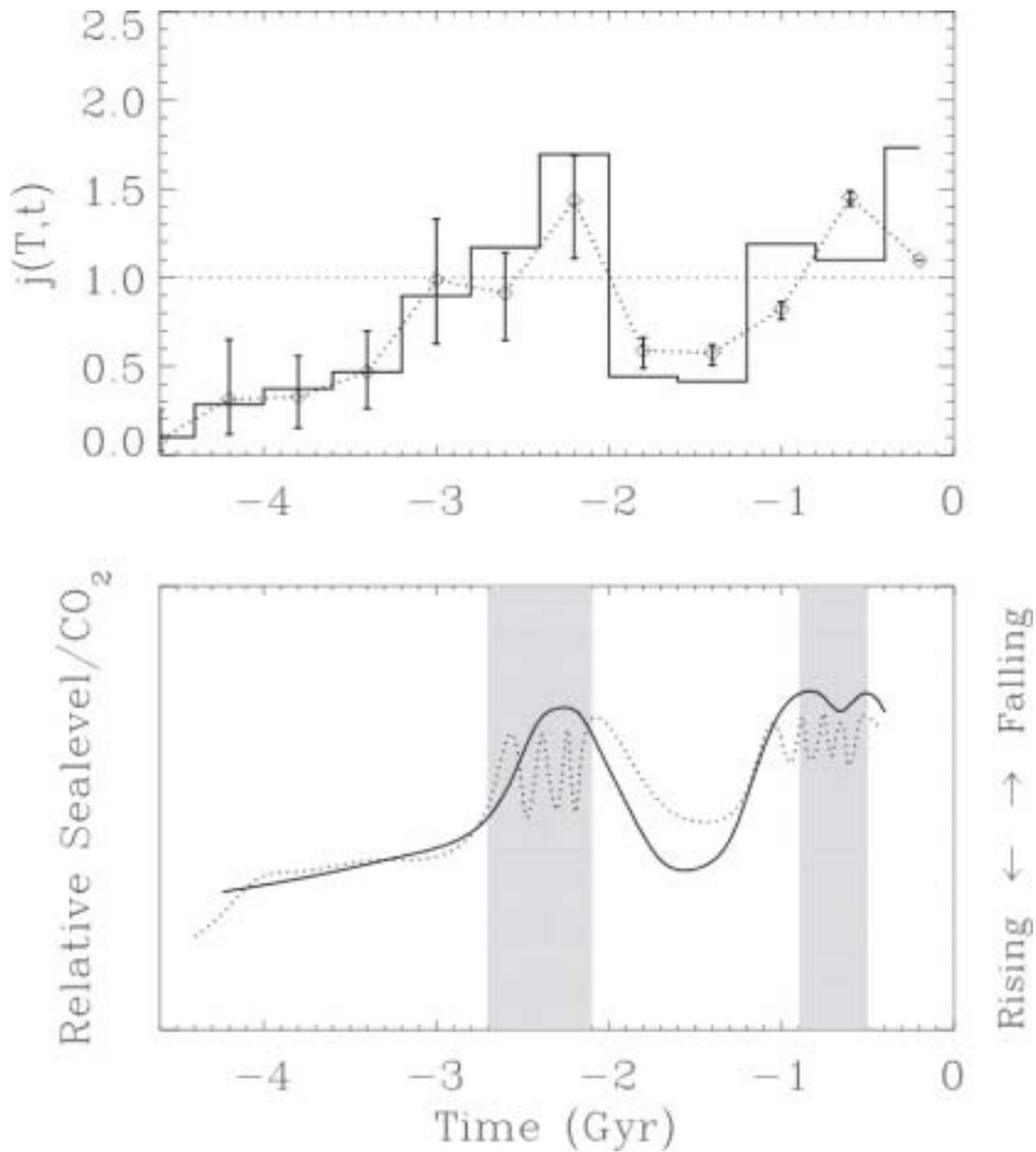

**Fig 4. Top panel: Variation of Galactic Cosmic Ray (GCR) particles with energy 10 GeV at 1 AU over the entire history of the solar system. The error bars indicate the modulation for particles of 5 GeV (lower bound) or 20 GeV, respectively (upper bound). Lower panel: Generalized evolution of Earth's climate represented by relative sea level changes (solid curve) and relative changes of atmospheric content of $CO_2$ (dashed curve). The extremes indicate either a Greenhouse climate (Warm) or an Icehouse (Cold), corresponding to**

**high sealevel/$CO_2$ level or low sealevel/$CO_2$ level respectively. The gray areas are known periods of glaciations.**